\documentclass{phbauth}
\usepackage{graphicx}

\begin{document}

\begin{frontmatter}

\title{Phase separation, percolation and giant isotope effect in
manganites}

\author{D.Khomskii\thanksref{thankdkhom}}

\address{Laboratory of Solid State Physics, Groningen University,
Nijenborgh~4, 9747~AG~Groningen, The~Netherlands}
\thanks[thankdkhom]{e-mail: khomskii@phys.rug.nl}

\begin{abstract}
Phase separation and a tendency to form inhomogeneous
structures seems to be a generic property of systems with strongly
correlated electrons. After shortly summarising the existing
theoretical results in this direction, I concentrate on the phenomena
in doped manganites. I discuss general theoretical results on the
phase separation at small doping and close to the doping $x=0.5$. The
``global'' phase diagram in this region is constructed. These
general results are illustrated on the example of the particular
system with rich and complicated
properties---(LaPr)$_{1-x}$Ca$_x$MnO$_3$ in which there exist a
ferromagnetic metallic (FM) phase and a charge ordered (CO)
insulating one. The experimental situation in this system is discussed
and the interpretation is given in the framework of the model with
competition of FM and CO, and the indications of phase separation
and percolative nature of this system are given. Giant isotope effect
observed in this situation is shortly discussed.
\end{abstract}

\end{frontmatter}
\let\<=\langle \let\>=\rangle \let\dg=\dagger \def\v#1{{\bf #1}}
The systems with correlated electrons present quite
specific class of compounds, many properties of which differ markedly
from those with ordinary band electrons.  Recently yet another
specific feature of these materials came to a forefront and attracts now
considerable attention---the often present in them tendency to phase
separation and creation of inhomogeneous states.  This tendency was
first studied theoretically in a number of models ($s$--$d$, or
double exchange
model~\cite{Nagaev,Kasuya,Dagotto,Kagan,Arovas}),
Hubbard model~\cite{Visscher}, $t$-$J$
model~\cite{Emery,Castellani} and was invoked to explain
properties of many real systems: magnetic semiconductors (see
e.g.~\cite{Nagaev2}), cuprates~\cite{Gor'kov} etc.

Depending on the specific situation the instability of homogeneous
state and the tendency to phase separation may result in a formation
of different structures: either random, percolation-like
networks~\cite{Babushkina,Gheong,Faeth} or regular
structures, e.g.\ stripes~\cite{Zaanen,Tranquada}.  Many
properties in this situation differ markedly from those of homogeneous
states, and it has to serve as a basis of the explanation of
experimentally observed phenomena in them.

There appear at present more and more indications that the formation
of inhomogeneous states with concomitant percolation behaviour is an
intrinsic feature of manganites with the colossal magnetoresistance
(CMR): percolation picture quite naturally explains many features of
manganites in a wide concentration range, and may even lie at the
core of the very phenomenon of
CMR~\cite{Gor'kov2,Littlewood}.  In this article I will give
a short summary of the theoretical situation with the phase separation
in manganites and of some of the experimental consequences and
evidences of it, based mostly on the experimental results of Moscow
groups (Babushkina, Balagurov, Fisher
et~al.)---see~\cite{Babushkina2,Babushkina3,Balagurov,Fisher}
and also other papers in these
Proceedings~\cite{Babushkina4,Belova}.  In particular, giant
isotope effect which is characteristic of this situation was studied
in details in these works and will be discussed shortly at the end of
this paper.

The electronic state of typical band-like systems like ordinary metals
or semiconductors is usually homogeneous.  This is to a large extent
caused by the Fermi pressure of electrons: the increase of the Fermi
energy with increasing electron density gives large positive contribution to the
bulk modulus of the system and thus stabilizes homogeneous state.

However if due to strong electron correlation the electrons become
localized (Mott--Hubbard localization), this positive contribution to
the bulk modulus disappears or is strongly reduced, and some other
factors may appear instead driving the system towards inhomogeneous
state.  As mentioned in the introduction, this instability of the
homogeneous state and the resulting tendency to phase separation seems
to be a generic feature of systems with strong electron correlation;
the homogeneous states are rather an exception than a rule, existing
formally only at rare isolated points of the phase diagram, such as
exactly half-filled case for Hubbard or $t$--$J$ model or
charge-ordered state $x=0.5$ in manganites (see below).

The traditional model applied for manganites is the double-exchange
model~\cite{Zener,deGennes}
\begin{eqnarray}
H=-t\sum_{\<ij\>}c^\dg_{i\sigma}c_{j\sigma}+J\sum_{\<ij\>}\v
S_i\v S_j&\nonumber\\
\hfill{}-J_H\sum\v S_ic^\dg_{i\sigma}\hat{\mbox{\boldmath$\sigma$}}
c_{i\sigma}\quad&
\end{eqnarray}
which describes conduction electrons
$c^\dg_{i\sigma}$, $c_{i\sigma}$ interacting by the Hund's rule
coupling $J_H$ with localized spins $\v S_i$ which by themselves
would form an antiferromagnetic state due to exchange
interaction~$J$.

Standard quasiclassical treatment of this situation~\cite{deGennes}
leads to the conclusion that for $J_H\gg t>J$ doping of the
insulating antiferromagnetic state (the increase of the concentration
of conduction electrons~$x$) leads to a gradual canting of
antiferromagnetic sublattices until at critical concentration
$x_c\sim J/t$ the system becomes ferromagnetic.  This is due to the
fact that for large $J_H$ the hopping of electrons is hindered by
the antiferromagnetic ordering, $t_{\it eff}=t\cos\frac\theta2$,
where $\theta$ is the angle between spins of the sublattices, and to
gain kinetic energy it is favourable to make the angle $\theta$
smaller. Simple calculations (minimization of the total energy
in~$\theta$) give for $J_H\gg t$ \ $\cos\frac\theta2=\frac tJ x$,
from which we get the condition for ferromagnetism induced by doping
($\theta=0$ for $x\ge x_c$).

However one immediately sees that in this treatment the homogeneous
canted state is absolutely unstable~\cite{Kagan,Arovas}: the
electron energy in this approximation is
\begin{equation}
E=E_0-\frac{t^2}{J}x^2
\end{equation}
so that $\partial^2 E/\partial x^2<0$, i.e.\ the compressibility of the system is
negative which signals the instability towards phase separation.  This
tendency survives in a more elaborate quantum
treatment~\cite{Kagan}, this is confirmed by the numerical
calculations~\cite{Dagotto}.

Thus the homogeneous canted state of double exchange systems is
absolutely unstable at low doping, and the system would phase separate
into regions of undoped antiferromagnet and region with higher
electron (or hole) concentration---ferromagnetic or strongly canted
metallic (FM) droplets.  These FM droplets may form percolative
network, which can explain transport and other properties of
manganites in this
region~\cite{Gor'kov2,Babushkina,Gheong}.

In reality other factors also play an important role in this
effect---notably the long-range Coulomb forces which oppose charge
segregation on a large scale.  Nevertheless the tendency to phase
separation survives even in this case~\cite{Emery,Nagaev2},
Coulomb interaction limiting the size of FM clusters.

\begin{figure}[t]
\begin{center}\leavevmode
\includegraphics[width=0.6\linewidth]{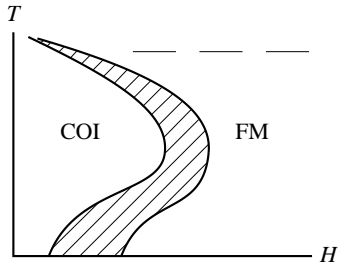}
\caption{Schematic form of the phase diagram in
Pr$_{1-x}$Ca$_x$MnO$_3$ for $0.3<x\leq0.45$ (by~\cite{Tokura})
}\label{fig1}\end{center}
\end{figure}

It is well established~\cite{Jirak,Tokura} that besides
antiferro- and ferromagnetic states, also charge-ordered states are
realized in manganites, especially at commensurate doping $x=0.5$ (one
electron or hole per two Mn's).  This state with simple checkerboard
charge ordering (CO) is an exact ground state for $x=0.5$, but often
it extends to other values of $x$, e.g.\ in
Pr$_{1-x}$Ca$_x$MnO$_3$---down to $x=0.3$~\cite{Jirak}.  For these
cases an interesting phase diagram was found in~\cite{Tokura}:
strong enough magnetic field renders system metallic, but with an
unusual reentrant behaviour in $(H,T)$-plane (see fig.~\ref{fig1})
(insulating CO state (COI) exists at intermediate temperature range,
but is transformed into FM state at lower temperatures; this
transition is accompanied by large hysteresis).

The first surprising feature of this phase diagram is negative slope
and reentrant character of the FM--COI transition line for~$x<0.5$.
The sequence of phases with temperature is determined by entropy: the
higher-temperature phase is always the one with larger entropy.  Here
however the CO (charge-{\em ordered\/}) state appears with {\em
increasing} temperature.

The answer is apparently that the low-tem\-pe\-ra\-tu\-re FM state is
not an ordinary disordered liquid, as compared to a CO ``crystal'',
but a {\em Fermi}-liquid with a unique ground state (Fermi-surface)
and consequently with zero entropy.  On the contrary, the CO
insulating state for $x<0.5$ is not fully ordered: as follows from
the neutron data~\cite{Jirak} the overall CO periodicity in this
state is the same as for $x=0.5$ (superlattice peaks at
$(\frac12,\frac12,\frac12)$) but one has to accommodate somewhere
the extra electrons (or Mn$^{3+}$ ions) present for $x<0.5$---and
this is apparently done in a random fashion which gives an extra
entropy of the CO state and drives the system into this state with
increasing temperature.

What is the exact nature of this partially ordered CO state?  There
are in general two options.  Either these extra electrons (Mn$^3+$)
are distributed randomly in one sublattice of the checkerboard CO
state typical for $x<0.5$, or there again exists phase separation,
with the electron-rich regions forming percolation---again
random!---network on a CO background.  One can give arguments that
most probably it is the latest situation which is realized in
practice.  Thus the situation close to doping $x=0.5$ resembles that
at small $x$: the system away from particular commensurate doping
tends to phase separate into a commensurate (here charge-ordered)
state and FM droplets.  As we will see below this picture permits to
explain many experimental results in the system
(La$_{1-y}$Pr$_y$)$_{1-x}$Ca$_x$MnO$_3$ and is confirmed by direct
experiments~\cite{Babushkina,Babushkina3,Balagurov,Fisher}.

Qualitative arguments supporting this conclusion can be drawn using
the pseudospin description of the charge ordering and corresponding
analogy with magnetic systems.  The simplest model describing the
formation of a CO state is that of spinless fermions with $nn$
repulsion
\begin{eqnarray}
H={}&-t\sum c^\dg_ic_j+V\sum_{\<ij\>}n_in_j-\mu\sum n_i,\label{eq3}\\
&\qquad n_i=c^\dg_ic_i\nonumber
\end{eqnarray}
where we inroduced also chemical potential $\mu$ to be able to
describe the situation with arbitrary doping.

One can easily show~\cite{Khomskii,Babushkina3} that for $V>2t$
the ground state of a system with one
electron per two sites ($n_{el}=x=\frac12$) is the checkerboard CO
state.  By introducing the pseudospin  variables $\sigma_i$ such that
$n_i=\frac12+\sigma_i^z$ ($\sigma^z=+\frac12$ corresponds to an
occupied site and $\sigma^z_i=-\frac12$ --- to an empty one) one can
model our system by the Hamiltonian
\begin{eqnarray}
H={}&&-t\sum(\sigma_i^\dg\sigma_j^-+{\rm
h.c.})\nonumber\\
&&{}+V\sum\sigma^z_i\sigma^z_j-\mu\sum\sigma_i^z+{\it
const}\,.
\label{eq4}
\end{eqnarray}
(Actually the model~(\ref{eq4}) is not exactly equivalent
to~(\ref{eq3}) due to different commutation relation of operators,
but the qualitative behaviour of both systems is very similar.)  The
CO state at $n=\frac12$ corresponds to an ``antiferro'' state with
total $\<\sigma^z\>=0$.

\begin{figure}[t]
\begin{center}\leavevmode
\includegraphics[width=0.6\linewidth]{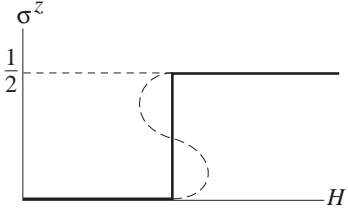}
\caption{Schematic behaviour of the total magnetization
$\<\sigma^z\>$ (in our case---electron density) at the metamagnetic
transition of Ising system
}
\label{fig2}\end{center}
\end{figure}

\begin{figure}[t]
\begin{center}\leavevmode
\includegraphics[width=0.6\linewidth]{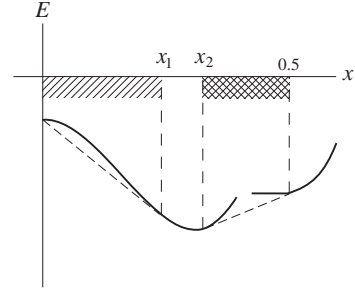}
\caption{The schematic global phase diagram of manganites for doping
$x<0.5$.  Curve~I is the energy of the homogeneous canted (De~Gennes)
state; curve~II --- the energy of the charge ordered state.  Maxwell
constructions are shown by dashed lines.  The hatched and
cross-hatched regions are regions of phase separation respectively
into pure antiferromagnetic / ferro.\ metal and charge ordered /
ferro.\ metal states.   The region between $x_1$ and $x_2$ is in
principle a homogeneous FM phase, although depending on the system
this region may shrink to zero.
}\label{fig3}\end{center}
\end{figure}

One can easily see that the situation with nonzero net
``magnetization'' $\<\sigma^z\>\neq0$ which corresponds to
$\<n\>\neq\frac12$ would describe the state with some of the
pseudospins of one sublattice reversed.  As is well known, the
increase of the ``magnetic field'' $\mu$ causes metamagnetic
transition---a jump of magnetization, which is actually the I~order
phase transition, see fig.~\ref{fig2}.  Consequently, the situation
with fixed net magnetization $0<\<\sigma^z\><\frac12$ (which
corresponds to a net electron density $\frac12<n<1$ or hole
concentration $x=1-n$ between $0$ and~$\frac12$) corresponds to
coexistence of two phases: ``antiferro'' (CO) one with
$\<\sigma^z\>=0$, and ``ferro'' one (all extra electrons forming one
big cluster with $\<\sigma^z\>=1$ or $n=1$, $x=0$). The vertical line
in fig.~\ref{fig3} may be viewed as a Maxwell construction with
unstable regions shown by dashed lines in fig.~\ref{fig3}. Thus these
qualitative arguments show that the system with doping close but
different from $x=\frac12$ may indeed phase separate into CO state
with $x=\frac12$ and another phase, presumably metallic, with some
$x<\frac12$.

One comes to essentially the same conclusions from quite different
point of view.  As recently noticed in~\cite{Soloviev}, with the
account of the CE-type magnetic structure of CO phase in manganites at
$x=0.5$~\cite{Jirak} the motion of charge carrier is essentially
one-dimensional, along the ferromagnetic zigzag chains in
CE-structure.  Corresponding band structure consists of bonding and
antibonding bands and nonbonding band in
between~\cite{Soloviev,vandenBrink}.  For $x=0.5$ the
bonding band is full and the other two bands are empty.  For $x>0.5$
the lower band becomes gradually depopulated, with corresponding
gradual change in band energy.  However the situation is highly
asymmetric around $x=0.5\,$: for $x<0.5$ (more electrons in the
system) it is the dispersionless nonbonding electron band which
becomes populated.  As a result the total band energy is constant for
certain range of concentrations $<0.5$ which again leads to the phase
separation.

The resulting ``global'' phase diagram of manganites for $x\leq0.5$
takes the form shown schematically in fig.~\ref{fig3}.  Maxwell
construction is shown there by dashed lines.  We see that there
exists two regions of phase separation: for $0<x<x_1$ the system
phase separates into undoped antiferromagnetic insulator ($x=0$) and
ferromagnetic metallic phase with $x=x_1$.  Close to half-filling the
system is again unstable and for $x_2<x<0.5$ it separates into the
mixture of the CO insulating antiferromagnetic phase
with~$x=0.5$, and a FM phase with~$x=x_2$.  The inclusion of Coulomb
interaction would modify the parameters of these phases and would
prevent full phase separation but rather stabilize ``fine-grained''
mixture of phases, but would not change the situation qualitatively.
The question whether these exists a ``window'' $x_1<x<x_2$ at which a
homogeneous FM state would be realized, is open: it may depend on
parameters of the specific system etc.  Thus e.g.\ such homogeneous
FM state may exist for La$_{1-x}$Ca$_x$MnO$_3$ (although it is
actually not known for certain at present); but such window is
definitely absent for Pr$_{1-x}$CaMnO$_3$~\cite{Jirak} which is
insulating for all~$x$.

The conclusions reached above show that the phase separation is a
generic property of manganites and probably also of many other
systems with correlated electrons.  What form the final state takes
depends on many specific details ignored in our general treatment,
such as specific type of magnetic structure, possible orbital
effects, Coulomb and electron--lattice interaction etc.  It may be
random percolation network with different lengthscales, or it may
even be a regular stripe-like structure.

Above we concentrated on the situation in manganites at $x\leq0.5$.
For overdoped (or electron-doped~\cite{Maignan}) manganites the
situation with respect to phase separation may be
similar~\cite{vdBrink2}, although the details may differ
significantly. In any case the treatment shortly presented
above~\cite{vandenBrink} definitely shows that in general the
situation for $x<0.5$ and $x>0.5$ is highly asymmetric---the
conclusion which agrees with all the experimental observations.

\begin{figure}[t]
\begin{center}\leavevmode
\includegraphics[width=0.6\linewidth]{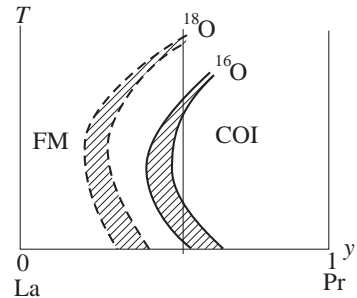}
\caption{Schematic form of phase diagram for the system
(La$_{1-y}$Pr$_y$)$_{0.7}$Ca$_{0.3}$MnO$_3$.  FM and COI are
ferromagnetic metallic and charge ordered insulating phases.  Shaded
regions with solid boundary is the phase transition region with
hysteresis for $^{16}$O-containing sample, and similar shaded
region with dashed boundary---that for $^{18}$O.
}\label{fig4}\end{center}
\end{figure}

In the last part of the paper I will shortly discuss experimental
evidences in favour of the theoretical picture described above.  One
of the most convenient systems to study these effects, especially the
interplay of CO insulating and FM states, is the system
(La$_{1-y}$Pr$_y$)$_{1-x}$Ca$_x$MnO$_3$ studied in details
in~\cite{Babushkina,Gheong,Babushkina2,Babushkina3,Balagurov,Fisher,Babushkina4,Belova}.
As already mentioned above, La$_{1-x}$Ca$_x$MnO$_3$ has a
ferromagnetic metallic state with the colossal magnetoresistance for
$0.16\leq x\leq0.5$. Pr$_{1-x}$Ca$_x$MnO$_3$ however is a
charge-ordered insulator in this concentration range.  Thus by
changing the ratio La/Pr in a mixed compound we can go from FM to
COI phases, even with fixed doping~$x$.  It turns out that close to
the crossover from one phase to another which is characterized by
strong hysteresis, all the properties very strongly depend on isotope
composition: one can even induce a metal--insulator transition by
only substituting $^{18}$O instead of
$^{16}$O~\cite{Babushkina2,Babushkina3,Zhao}.  The
schematic phase diagram for this system which follows from all these
experiments has the form shown in fig.~\ref{fig4} where the borderline
between FM and COI phases is marked for both $^{16}$O and $^{18}$O
systems.

Quite characteristic for this case, besides the very specific form of
the phase diagram, with the reentrant behaviour and with giant
isotope effect, are many properties which signal an intrinsic
inhomogenuity in this situation. Many properties in this regime find
natural explanation in the percolation picture.  Such are e.g.\ the
magnetic properties~\cite{Babushkina4}, strongly nonlinear $I$--$V$
characteristics of these samples~\cite{Babushkina}, many other
transport properties~\cite{Gheong}.  The inhomogeneous nature of
these samples was visualized directly in~\cite{Gheong}.

\begin{figure}[t]
\begin{center}\leavevmode
\includegraphics[width=0.6\linewidth]{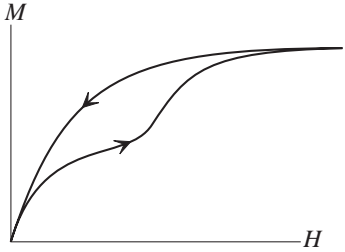}
\caption{The schematic dependence of the magnetization on magnetic
field in (La$_{1-y}$Pr$_y$)$_{0.7}$Ca$_{0.3}$MnO$_3$
(by~\cite{Fisher})
}\label{fig5}\end{center}
\end{figure}

I will discuss here only one extra spectacular effect which finds
natural explanation in the picture of phase separation.  The
behaviour of the magnetization of these systems with external field
shows very unusual ``shifted hysteresis'' behaviour shown in
fig.~\ref{fig5}~\cite{Belova}.  $M(H)$ curves are symmetrical for
$H<0$. Similar but less spectacular behaviour was also seen
in~\cite{Amaral,Coey}.
Simultaneously with this unusual $M(H)$ curves the magnetization
shows marked relaxation behaviour, with characteristic times reaching
tens of minutes.

The most natural explanation of all these facts can be obtained in a
picture of inhomogeneous state, with ferromagnetic droplets immersed
in the charge ordered antiferromagnetic matrix (neutron studies
directly confirm this interpretation~\cite{Balagurov}).  The first
part of $M(H)$ curves is explained in this picture by the
conventional orientation of the moments in ferromagnetic regions
(an ordinary hysteresis close to $H=0$ is also seen but apparently it
is a very soft magnet, and the coercive force is very low).

Starting from the fields $\sim1$--$2\,$T the ferromagnetic regions
start to grow, ``eating up'' the antiferromagnetic ones, until at
$H\simeq4$--$5\,$T they occupy the whole sample, and the
magnetization reaches saturation.

At the reverse change of field there exists marked hysteresis, seen
also in resistivity and in other properties.  One can successfully
model this behaviour, including also time dependence, by assuming the
existence of two locally stable states, FM and CO antiferromagnetic
one, with certain energy barriers between them.  Thus this
experiment gives yet another clear indication of an inhomogeneous
state at least of this particular system but most probably
present also in many other situations.

The last comment concerns the nature of giant isotope effect in this
situation.  As follows from the theoretical considerations based on
model~(\ref{eq3})~\cite{Khomskii,Babushkina3} the balance
between COI and FM state strongly depends on the ratio of the
electron (or hole) bandwidth $\sim t$ and effective $nn$ repulsion
$V$: CO state exists only for $V>V_c\sim2t$.  By changing isotope
composition we in principle modify electron hopping $t$, either
because of the polaron band narrowing
\begin{equation}
t\;\longrightarrow\;t^*=t\exp\left(-\frac{E_{\it
pol.}}{\omega}\right)
\end{equation}
where $E_{\it pol.}$ is the polaron binding energy and $\omega$ is
the typical phonon frequency; or, in the absence of polaron effects,
the effective hopping decreases simply due to the averaging of $t(r)$
over lattice vibrations (both thermal and zero-pont
ones)~\cite{Babushkina3}.  Both these effects work in the right
direction, reducing the bandwidth for heavier $^{18}$O ions and
shifting the equilibrium in the direction of the CO state.  However the
estimates show that the changes of $t_{\it eff}$ are still rather
small, and it is not completely clear what is the real nature of this
strong isotope dependence.  (Actually this problem is the same as the
one encountered in the whole class of manganites where relatively
small changes in crystal structure caused e.g.\ by going from
La$_{1-x}$Ca$_x$MnO$_3$ to Pr$_{1-x}$Ca$_x$MnO$_3$ change all the
properties drastically.)  One possible explanation of this strong
sensitivity of the behaviour of e.g.\
(La$_{1-x}$Pr$_x$)$_{0.7}$Ca$_{0.3}$MnO$_3$ to oxygen isotope
composition~\cite{Babushkina2,Babushkina3,Balagurov,Fisher,Babushkina4,Belova}
may again lie in the behaviour close to percolation threshold where as
is well known relatively minor change of parameters may lead to a
drastic modification of the properties of the system.

In conclusion, we presented above theoretical arguments which show
that the instability towards phase separation and formation of
inhomogeneous states is an intrinsic tendency of many systems with
correlated electrons, in particular CMR manganites.  Such phase
separation should exist both at low doping level and close to
$x\simeq0.5$.  Whether the optimally doped phase with CMR is
homogeneous or it is also inhomogeneous at small scale (with possible
dynamic phase separation) is not clear at present.  In any case the
tendency to phase separation and the resulting percolation picture
should be taken into account in the interpretation of many
experiments in manganites.  In many cases this is the most natural
and possibly the only explanation of experimental observations.  Some
of them were shortly discussed or mentioned in our paper; there exist
now many other indications of the same phenomenon which we had no
chance to discuss (e.g.\ neutron~\cite{Hennion} or NMR~\cite{Allodi}
results). All these data, together with the theoretical arguments
presented above, point to the crucial role played by the phase
separation and percolation in the physics of CMR manganites and
presumably in many other systems with correlated electrons.

\begin{ack}
I am very grateful to many coworkers and colleagues with
whom I extensively discussed the problems described above.  This work
was supported by the Netherlands Foundation of the Fundamental Study
of Matter (FOM), by the European Network OXSEN and by the grant
INTAS-97-0963.
\end{ack}

\end{document}